\def\bbox{\vec}
\def\institute#1{\gdef\@institute{#1}}
\def\@institute{\inst{1} First address\\
                   \inst{2} Second address}
\def\inst#1{$(^{#1})$}
\newtoks\ACK \ACK={Acknowledgments}
\def\ack#1{\section*{\the\ACK}#1}
\def\review#1, #2, 1#3#4#5, #6 {{\sl#1\/} {\bf#2} (1#3#4#5) #6}
\def\book#1, #2, #3, 1#4#5#6, #7 {#1 (#2, #3, 1#4#5#6) p. #7}

\documentstyle[preprint,aps] {revtex}

\begin{document}

\title{Light Front Treatment of 
 Nuclei and Deep Inelastic Scattering}
\author{G. A. Miller\vspace{15pt}}
\address {permanent address Department of Physics, 
Box 351560, \\
University of Washington, Seattle, WA 98195-1560
\\
Stanford Linear Accelerator Center, Stanford University,\\
Stanford, California 94309   
\\
national Institute for Nuclear Theory, Box 35150, \\
University of Washington, Seattle, WA 98195-1560
}
\maketitle
\section{INTRODUCTION}
A light front treatment of the nuclear wave function is developed and
 applied, using the mean field approximation, to infinite nuclear
 matter.  The nuclear mesons are shown to carry about a third of the
 nuclear plus momentum $p^+$; but their momentum distribution has 
 support only at $p^+=0$, and the mesons do not contribute to nuclear
 deep inelastic scattering.  This zero mode effect occurs because the
 meson fields are independent of space-time position.
\section{DISCUSSION}
The discovery  that the  deep inelastic scattering  
 structure function of a  bound nucleon differs from that of a free one 
(the EMC effect\cite{EMCrefs}) 
changed the way that physicists viewed the nucleus. With a 
 principal effect that
  the plus  momentum (energy plus third component of the 
momentum, $p^0+p^3\equiv p^+$) carried by the  valence quarks  is 
less for a bound nucleon than for a free one, 
quark and nuclear physics could not be viewed as being 
 independent.  

The interpretation of the experiments requires that the role of conventional 
effects, such as nuclear binding, be assessed and  understood\cite{EMCrevs,sfrel}. 
Nuclear binding is supposed to be relevant because the plus momentum 
of a bound nucleon is reduced 
by the binding energy, and  so is that of  its  confined  quarks.
Conservation of momentum implies that if nucleons lose momentum,
other constituents such as nuclear pions\cite{ET}, 
 must gain momentum. This partitioning of the total plus momentum amongst 
the various constituents is the   momentum sum rule.
 Pions  are quark anti-quark pairs so that a specific enhancement of  
the nuclear antiquark momentum distribution
  is  a  testable \cite{dyth} consequence of this idea. 
A nuclear Drell Yan experiment \cite{dyexp}, 
was performed, but  no influence  of nuclear pion enhancement was seen.
This led
Ref.~\cite{missing} 
to  question the idea of the pion as a dominant carrier of the 
nuclear force.
\section{NUCLEAR CALCULATION}
Here  a closer look at the relevant nuclear theory is taken,  and 
the momentum sum rule is studied. This talk  is based on Ref.\cite{me}.

 The structure function depends on the Bjorken
variable 
$x_{Bj}$ which in the parton model 
 is the ratio of the quark plus momentum to that of the target. Thus $x_{Bj}
=p^+/k^+$, where  $k^+$ is the 
plus momentum of a nucleon bound in the nucleus, so
a more direct relationship between the necessary 
nuclear  theory and experiment occurs by
using a theory in which $k^+$ is one of the canonical variables.
Since $k^+$ is conjugate to a spatial variable  $x^-\equiv t-z$, 
it is natural to quantize the dynamical variables 
at the equal light cone time variable of $x^+\equiv t+z$. 
To use such a formalism is to use  light front quantization, 
which requires a new derivation of  
the nuclear wave function,  because previous work used
  the equal time formalism.

Such a derivation is provided here, using a simple 
model in which the 
nuclear constituents are nucleons $\psi$ (or $\psi')$,
 scalar mesons $\phi$\cite{scalar} and vector mesons
$V^\mu$.
The  Lagrangian ${\cal L}$ is given by 
\begin{eqnarray}
{\cal L} ={1\over 2} (\partial_\mu \phi \partial^\mu \phi-m_s^2\phi^2) 
-{1\over  4} V^{\mu\nu}V_{\mu\nu} +{m_v^2\over 2}V^\mu V_\mu 
\nonumber\\
+\bar{\psi}^\prime\left(\gamma^\mu(i \partial_\mu-g_v\;V_\mu) -
M -g_s\phi\right)\psi' \label{lag}
\end{eqnarray}
where the bare masses of the nucleon, scalar and vector mesons are given by 
$M, m_s,$  $m_v$, and  $V^{\mu\nu}=
\partial ^\mu V^\nu-\partial^\nu V^\mu$. This Lagrangian may be thought of 
as a low energy effective 
theory for nuclei under normal conditions.
 
This hadronic model, when evaluated in mean field approximation, 
gives\cite{bsjdw}  at least a qualitatively 
good description of many (but not all) 
nuclear properties and reactions. The aim here is
to use a simple Lagrangian to study the effects that one might obtain by using
a light front formulation. It is useful to
simplify this first calculation by studying infinite nuclear matter.

The light front quantization procedure  necessary
to treat  nucleon interactions with scalar and vector mesons
was derived by Soper\cite{des71} and also by
Yan and collaborators\cite{yan12,yan34}. Glazek and
Shakin\cite{gs} used  a  Lagrangian 
containing nucleons and scalar mesons  to study  infinite nuclear matter. 
Here both vector and scalar mesons are included, and the 
nuclear plus momentum  distribution is  obtained.

Next examine the field equations. The 
nucleons satisfy
\begin{equation}
\gamma\cdot(i\partial-g_v V)\psi'=(m+g_s\phi)\psi'.
\label{dirac}
\end{equation}
The number of independent degrees of freedom for light front 
field theories is fewer than in the usual theory. One defines 
projection operators $\Lambda_\pm\equiv \gamma^0\gamma^\pm/2$ 
and the independent Fermion degree of freedom is 
$\psi'_+=\Lambda_+\psi'$. 
The  relation between
$\psi'_-$ and $\psi'_+$
is very complicated unless one may set the plus component of the vector field
to zero. This is a matter of a choice of gauge for 
QED and QCD, but the non-zero mass of the vector meson prevents such a choice 
here. Instead, one simplifies the equation
for $\psi'_-$ by\cite{yan34} 
   transforming  the Fermion field according to 
$\psi'=e^{-ig_v\Lambda(x)}\psi $ with $\partial^+ \Lambda=V^+$ which 
yields
\begin{eqnarray}
(i\partial^--g_v \bar V^-)\psi_+=(\bbox{\alpha}_\perp\cdot 
(\bbox{p}_\perp-g_v\bbox{\bar V}_\perp)+\beta(M+g_s\phi))\psi_-\nonumber\\
i\partial^+\psi_-=(\bbox{\alpha}_\perp\cdot 
(\bbox{p}_\perp-g_v\bbox{\bar V}_\perp)+\beta(M+g_s\phi))\psi_+\, \label{yan}
\end{eqnarray}
where
\begin {equation}
 \partial^+\bar V^\mu=\partial^+V^\mu-\partial^\mu V^+\label{vbar}
\end{equation}

The field equations for the mesons are 
\begin{eqnarray}
\partial_\mu V^{\mu\nu}+m_v^2 V^\nu=g_v\bar \psi\gamma^\nu\psi\nonumber\\
\partial_\mu \partial^\mu\phi+m_s^2\phi=-g_s\bar\psi\psi.
\end{eqnarray}

We now introduce the mean field approximation\cite{bsjdw}. The coupling
 constants are  
considered strong and the Fermion density  large. Then the meson 
fields can be approximated as classical-  the 
sources of the meson fields  are replaced
 by their expectation values. In this case, the nucleon mode functions will
be plane waves and the nuclear matter ground state can be 
 assumed to be a  normal Fermi gas,  of Fermi momentum
$k_F$, and  of large volume $\Omega$ in its rest frame. 
We consider the case that there is an equal number of protons and neutrons.
Then the meson fields are constants given by 
\begin{eqnarray}
\phi=-{g_s\over m_s^2} \langle \bar \psi \psi\rangle\nonumber\\
V^\mu={g_v\over m_v^2} \langle \bar \psi
\gamma^\mu\psi\rangle=\delta^{0,\mu}{g_v\rho_B\over m_v^2},\label{mfa}
\end{eqnarray}
where $\rho_B=2k_F^3/3\pi^2$.
This result that $V^\mu$ is a constant, along with Eqs.~(\ref{vbar}) and 
(\ref{mfa}),  
 tells us that the only non-vanishing component of $\bar V$ is 
$\bar {V}^-= V^0$. 
The expectation values refer to the nuclear matter ground state.
 
With this, the light front Schroedinger 
equation for the modes of the field operator $\sim e^{ik\cdot x}$
and can be obtained from Eq.~(\ref{yan}) as\cite{exp}
\begin{equation}
(i\partial^--g_v \bar V^-)\psi_+
={\bbox {k}_\perp^2 +(M+g_s\phi)^2\over k^+}\psi_+. \label{sol}
\end{equation}
The light front eigenenergy $(i\partial^-\equiv k^-)$
is the sum of a kinetic energy term in which the mass is shifted by the
presence of the scalar field, and an constant 
energy arising from the vector field.
Thus the nucleons 
have a  mass $M+g_s\phi$ and move in plane wave states. The nucleon 
field operator is constructed using the solutions of 
Eq.~(\ref{sol}) as the plane wave basis states. This means that 
the nuclear matter ground state, defined by operators that create and 
destroy baryons in eigenstates of Eq.~(\ref{sol}), is the correct 
wave function and that Equations~ (\ref{mfa}) and (\ref{sol})
represent the solution of the approximate
field equations, and the diagonalization of the  Hamiltonian. 

The computation  of the energy and plus 
momentum distribution proceeds from taking the appropriate expectation
values of 
$T^{\mu\nu}$:
\begin{equation}
P^\mu={1\over 2}\int d^2 x_\perp dx^- \langle T^{+\mu}\rangle.\label{pmu}
\end{equation}
We are concerned with the light front energy $P^-$ and momentum $P^+$.
Within the mean field approximation one finds
\begin{eqnarray}
T^{+-}= m_s^2\phi^2 +2\psi_+^\dagger (i\partial^--g_v\bar V^-)\psi_+
\nonumber\\
T^{++}=m_v^2 V_0^2+ 2\psi^\dagger_+i\partial^+\psi_+.
\end{eqnarray}
 Taking the nuclear matter expectation
value of $T^{+-}$ and $T^{++}$ and performing the spatial integral of 
Eq. (\ref{pmu}) leads to the  result 
\begin{eqnarray}
{P^{-}\over \Omega}&=& m_s^2
\phi^2 +{4\over (2\pi)^3}\int_F d^2k_\perp dk^+ {\bbox{k}_\perp^2+ 
(M+g_s\phi)^2\over k^+}\label{pminus}\\
{P^{+}\over \Omega}&=& m_v^2 
V_0^2 +{4\over (2\pi)^3}\int_F d^2k_\perp dk^+ k^+.\label{pplus}
\end {eqnarray}
The subscript F denotes that $\mid\vec k\mid<k_F$   with $k^3$ defined
by the relation 
\begin{equation}
k^+=\sqrt{(M+g_s\phi)^2+\vec k^2}+k^3.\label{kplus}
\end{equation}

The expression for the energy of the system $E={1\over 2}(P^++P^-)$\cite{gs}, 
is the same as in the usual treatment\cite{bsjdw}. This can be seen by
summing equations (\ref{pminus}) and (\ref{pplus}) and changing 
integration variables using ${dk^+\over k^+}={dk^3\over 
\sqrt{(M+g_s\phi)^2+\vec k^2}}$. This equality 
of energies is a nice check on the present result because  a manifestly  
covariant solution
 of the present problem, with the usual energy,  has been obtained\cite{sf}.
 Moreover,
setting ${\partial E\over \partial \phi}$ to zero reproduces the field
equation for $\phi$, as is also usual. 
Rotational invariance, here the relation
$P^+=P^-$, follows as the result of minimizing the energy per particle at
fixed volume with respect to $k_F$,
or minimizing the energy with respect to the volume\cite{gs}.
The parameters $g_v^2M^2/m_v^2=195.9$ and  $g_s^2M^2/m_s^2=267.1$ 
have been chosen
\cite{chin} so as to give the binding energy per particle of nuclear matter
as 15.75 MeV with $k_F$=1.42 Fm$^{-1}$. In this case, solving the 
equation for $\phi$ gives 
 $M+g_s\phi=0.56\;M$.

\section{NUCLEAR PLUS MOMENTUM DISTRIBUTIONS}
The use of Eq.~(\ref{pplus}) and these parameters leads immediately to the
result that only 65\% of the nuclear plus momentum is carried by the nucleons;
the remainder is carried by the mesons. This is a much smaller fraction than
is found in typical nuclear binding models\cite{EMCrevs,sfrel}.
 The nucleonic momentum distribution which 
is the input to calculations of the nuclear structure function of
primary  interest here. This function can be 
computed from the integrand of Eq.(\ref{pplus}). The probability that
a nucleon has plus momentum $k^+$ is determined from the condition that
the plus momentum carried by  nucleons, $P^+_N$, is given by 
$P^+_N/A=\int
dk^+\;k^+ f(k^+)$, where $A=\rho_B\Omega$.
 It is convenient to 
use the  dimensionless variable $y\equiv {k^+\over \bar{M}}$
with $\bar{M}=M-15.75 $ MeV. Then Eq.(\ref{pplus}) and simple algebra leads to
the equation
\begin{equation}
f(y)={3\over 4} {\bar{M}^3\over k_F^3}\theta(y^+-y)\theta(y-y^-)\left[
{k_f^2\over \bar{M}^2}-({E_f\over \bar{M}}-y)^2\right],
\end{equation}
where
$y^\pm\equiv {E_F\pm k_F\over \bar{M}}$ and
$E_F\equiv\sqrt{k_F^2+(M+g_s\phi)^2}$. 
Similarly the baryon number distribution $f_B(y)$
 (number of baryons per $y$, normalized to unity) can be
determined from the expectation value of $\psi^\dagger\psi$. The result is
\begin{eqnarray}
f_B(y)={3\over 8} {\bar{M}^3\over k_F^3}\theta(y^+-y)\theta(y-y^-)
\nonumber\\
\left[ (1+{E_F^2\over \bar{M}^2y^2})(
{k_f^2\over \bar{M}^2}-({E_F\over \bar{M}}-y)^2)
-{1\over 2y^2}({k_F^4\over \bar{M}^4}-({E_F\over \bar{M}}-y)^4)
\right],
\end{eqnarray}
which is different than 
$f(y)$.

The nuclear deep inelastic structure function, $F_{2A}$
can be obtained from the light front distribution function
$f(y)$ and the nucleon structure function
$F_{2N}$ using  the relation\cite{sfrel}
\begin{equation}
{F_{2A}(x)\over A}=\int dy f(y) F_{2N}(x/y), \label{deep}
\end{equation}
where $x$ is the Bjorken variable computed using the
 nuclear mass divided by $A$ ($\bar M$):  $x=Q^2/2\bar M \nu$.
This formula is the expression of the convolution model in which 
one means to assess, via  $f(y)$,  only  the influence of nuclear binding.
Consider the present effect of having the average value of 
$y$ equal to 0.65. Frankfurt and Strikman\cite{sfrel} use 
Eq.~(\ref{deep}) to argue 
that an average of 0.95 is sufficient to explain the 15\% 
depletion effect observed for the Fe nucleus.  
 One may also compare the 0.65 fraction with the result 
0.91 computed\cite{gmad}
for nuclear matter, including the effects of correlations, 
using equal time quantization. The present result then represents 
a very strong binding effect, even though this infinite nuclear matter 
result can not be compared directly with the experiments using Fe 
targets. One might think that the mesons, which cause this binding, 
would also have huge effects on deep inelastic scattering.

It is necessary to 
determine the  momentum distributions of the  mesons.
The mesons contribute 0.35 of the total 
nuclear plus momentum, but we need to 
know how this is distributed over different individual values.
The paramount feature is 
that $\phi$ and $V^\mu$ are the same constants for any  and all 
values of the spatial coordinates $x^-,\bbox{x}_\perp$. This means that
the related momentum distribution can only be proportional to a delta function
setting both the plus and $\perp$ components of the momentum to zero.
This result is attributed to the mean field approximation, in which the meson
fields are treated as classical quantitates. Thus the finite plus momentum 
can be thought of as coming from an infinite number of quanta, each carrying
an infinitesimal amount of plus  momentum. A plus momentum of 0 can only be
 accessed experimentally at $x_{Bj}=0$, which requires an infinite amount 
of energy. Thus, in the mean field approximation,
 the scalar and vector mesons can not contribute to deep 
inelastic scattering. The usual term for a field that is constant over space
is a zero mode, and  the present Lagrangian provides a simple example.
For finite nuclei, the mesons would carry a very small momentum of scale given
by the inverse of the nuclear radius, under the mean field approximation. 
If fluctuations were to be included, the relevant momentum scale would be  
of the order of the inverse of the average distance between
nucleons (about 2 Fm).

\section{SUMMARY AND ASSESSMENT}
The Lagrangian of Eq.~(\ref{lag}) and its evaluation in mean 
field approximation for nuclear matter have been 
used to provide a simple but semi-realistic example. 
It would be premature to compare the present results with data.
The  specific numerical results of the present work
are far less relevant than the emergent central
feature that the mesons responsible for nuclear 
binding need not be accessible in  deep inelastic scattering. 
Another interesting feature is that $f(y)$ and $f_B(y)$ are not 
the same functions.

More generally, we view the present model as being one of 
a class of models in which the mean field plays an important role.
For such models nuclei would have constituents that contribute to the 
momentum sum rule but do not contribute to deep inelastic scattering. 
Thus the predictive and interpretive power of the momentum sum rule is vitiated.
%
%
\ack{
This work is partially supported by the USDOE. I thank the SLAC theory group
 and the national INT for their hospitality.
I  thank S.J. Brodsky, L. Frankfurt, 
 S. Glazek, C.M. Shakin and M. Strikman for useful discussions.}
%

%

\begin{thebibliography}{100}
\bibitem{EMCrefs}  Aubert J., {\it et al.,} \review Phys. Lett.,  B123, 
1982,  275.
\bibitem{EMCrevs} Arneodo M., \review Phys. Rep., 240, 1994, 301.
\bibitem{sfrel} Frankfurt L.L., Strikman M.I., \review Phys. Rep.,  
160, 1988,  235.
\bibitem{ET} Ericson M., Thomas A.W., \review Phys. Lett., 
B128, 1983, 112.
\bibitem{dyth} Bickerstaff R.P., Birse M.C., Miller G.A., 
\review Phys. Rev. Lett., 53, 1984, 2532.
\bibitem{dyexp} Alde D.M., et al. \review Phys. Rev. Lett.,  64, 1990, 2479.
\bibitem{missing} Bertsch G.F., Frankfurt L., Strikman M., 
\review Science,  259, 1993, 773. 
\bibitem{me} Miller G.A., 1997 preprint, nucl-th/9702036,  
Phys. Rev. C. July 1997
\bibitem{notation} Our notation is that a four vector $A^\mu$ is defined by
the plus, minus and perpendicular components as
$(A^0+A^3,A^0-A^3,\bbox{A}_\perp)$.
\bibitem{scalar} The scalar mesons are meant to represent the two pion 
exchange potential which causes much of the medium range attraction 
between nucleons, as well as the effects of a fundamental scalar meson. 
Thus the pion is an important implicit part of the present Lagrangian.
\bibitem{bsjdw} Serot B.D., Walecka J.D., \review Adv. Nucl. Phys., 
16, 1986, 1; IU-NTC-96-17, Jan. 1997 nucl-th/9701058
\bibitem{exp} The symbol for the nucleon field operator and the mode functions
of that field is taken to be the same -$\psi$ to reduce the amount 
of notation.
\bibitem{des71} D.E. Soper, ``Field Theories in the Infinite Momentum Frame'',
SLAC pub-137 (1971); 
D.E. Soper, Phys. Rev D4, 1620 (1971); J.B. Kogut and D.E. Soper,
Phys. Rev. {\bf D1}, 2901 (1971).
\bibitem{yan12} Chang S-J., Root R.G., Yan T-M., \review Phys. Rev., 
D7, 1973, 1133; {\it ibid} (1973) 1147. 
\bibitem{yan34} Yan T-M., \review Phys. Rev., D7, 1974, 1760; 
{\it ibid} (1974) 1780. 
\bibitem{gs} Glazek St., Shakin C.M., \review Phys. Rev., C44, 1991, 1012.
\bibitem{sf} Serot B.D., Furnstahl R.J., \review Phys. Rev., C43, 1991, 105.
\bibitem{chin} Chin S.A., Walecka J.D., \review Phys. Lett., B52, 1974, 24.
\bibitem{gmad} Dieperink A.E., Miller G.A., \review Phys. Rev.,
  C44, 1991, 866.
%
\end{thebibliography}
\end{document}